# SIMULAÇÃO DA DINÂMICA DO USO DO SOLO EM PARAGOMINAS-PA: DIFERENÇAS NAS REGRAS ESPACIAIS ENTRE ÁREAS DE ASSENTAMENTOS E AGROPECUÁRIA COMERCIAL

**Reinis Osis**
Université du Maine / reinis.osis.etu@univ-lemans.fr
**François Laurent**
Université du Maine / francois.laurent@univ-lemans.fr
**René Poccard-Chapuis**
Cirad - UMR SELMET / renepoccard@gmail.com

**Eixo Temático: 1** Desenvolvimento Rural Sustentável, Territorialidade e Políticas Públicas

**Resumo**

O objetivo deste trabalho é apresentar os resultados de simulações de dinâmica de uso do solo no município de Paragominas-PA. Estas simulações foram baseadas em modelos construídos a partir de dados de uso do solo e variáveis espaciais do meio natural e de infraestrutura. Duas unidades espaciais foram analisadas: a zona central de produção agropecuária comercial e a zona de assentamentos e propriedades pequenas a Leste do município. Os resultados mostram dinâmicas espaciais distintas entre as zonas analisadas, dentre as quais podemos destacar o papel das características do solo, em que associadas à topografia e ao histórico de ocupação, fazem parte do contexto em que é definida a racionalidade dos produtores. Considerando a transição de floresta para pastagem, na zona de agricultura comercial os solos mais frequentemente associados a esta são os arenosos. Isso levanta a seguinte hipótese: os desmatamentos que ocorreram no período estão relacionados às atividades pecuárias. A pecuária privilegia o acesso à água e a baixa fertilidade das areias não afeta fortemente a produção. Por outro lado, a expansão da soja se deu preferencialmente sobre pastos já existentes e sobre solos argilosos (Argila de Belterra), reduzindo assim disponibilidade de pastos sobre estes tipos de solos. A importância relativa dos tipos de solo aumenta com o tempo, indicando a valorização de terras de acordo com a qualidade do solo. Por outro lado, na zona de assentamentos, a transição de floresta para pastagens e cultivos familiares se deu preferencialmente sobre a Argila variegada. Porém, é possível que a predominância desta transição sobre esta textura tenha se dado devido principalmente à trajetória de ocupação desta zona. Historicamente os fundos dos vales arenosos foram os primeiros a serem ocupados, sendo que a continuidade do desmatamento se deu em direção às encostas onde predomina a Argila variegada e às chapadas com Argila de Belterra. Estas associações observadas indicam que, dentro de um contexto mais amplo de fatores sociais, econômicos e políticos, os fatores naturais variáveis no espaço são importantes para a escolha dos manejos nas propriedades, mas que estes se fazem de maneira distinta no território, e o melhor conhecimento destas relações são úteis para o planejamento territorial.

**Palavras-Chave:** Dinâmica do uso do solo, Simulação, Amazônia, Solo.

**Abstract**

The aim of this paper is to present the results of the land use dynamic simulations in the municipality of Paragominas-PA. The simulation is based on models built from past land use data and spatial variables of the natural environment and infrastructure. Two spatial units were analyzed: the central area of commercial agricultural production and the area of settlements and smallholdings the east. The results show distinct spatial dynamics between the analyzed areas, among which we highlight the role of soil characteristics and, associated with the topography and the occupation history, are part of the context in which is defined rationality producers. Considering the transition from forest to pasture in the commercial farming area most often associated soils are sandy. This raises the following hypothesis: the deforestation that occurred in the period are related to livestock activities. Livestock favors access to water and low fertility sands does not affect production. On the other hand, the soybean expansion occurred preferentially on existing pastures and on clay soils (Belterra clay), reducing the availability of pastures on these soils. The relative importance of types of soil increases with time. In the area of settlements, the transition from forest to pasture and family crops occurred preferentially on the variegated clay. However, it is possible that the prevalence of this transition on this texture has been given due more to the history of occupation of this area. Historically the sandy valleys were the first to be occupied, and the continuity of the deforestation occurred toward the slopes dominated by variegated clay and plateaus with Belterra clay. These associations observed indicate that, within a wider context of social, economic and political factors, natural variable factors in space are important for the choice of managements in the properties, but they are done differently in the territory, and the best knowledge of these relationships are useful for territorial planning.

**Keywords:** Land use dynamics, Simulation, Amazon, Soil.

**1. Introdução**

Para analisar a dinâmica espacial do uso do solo, modelos de simulação dinâmica têm sido desenvolvidos e aplicados para a Amazônia, tanto para compreender a dinâmica passada quanto para projetar cenários futuros da paisagem, o que se revela de grande utilidade para o planejamento territorial (PERZ et al., 2009).

Nesse sentido, o presente trabalho pretende analisar a dinâmica do uso do solo no contexto de um território de fronteira agrícola consolidada na Amazônia Oriental, especificamente o município de Paragominas, conhecido pelo histórico passado de altas taxas de desmatamento e uma recente transformação política e social originada localmente no sentido de reduzir o desmatamento (PINTO et al., 2009).

A recente maior restrição da expansão horizontal das propriedades, sobretudo a partir de 2004, coloca os produtores em um novo contexto, sendo induzidos a novas organizações no espaço. No entanto, diferentes unidades de paisagem podem ser identificadas no município e cada uma pode apresentar uma dinâmica espacial própria.

A hipótese norteadora do trabalho considera que as características dos recursos naturais variáveis no espaço e o contexto socioeconômico, político e fundiário influenciam a racionalidade dos produtores no que se refere à lógica de ocupação do espaço, e a compreensão desta lógica permite elaborar modelos de dinâmica de uso do solo. A análise das mudanças de uso do solo entre 2014 e 2013 e como estas mudanças se relacionam com variáveis espaciais proporciona um reflexo da racionalidade dos produtores que é influenciada por estas variáveis.

O presente trabalho, portanto, pretende apresentar os resultados preliminares da análise das regras espaciais que comandam a dinâmica de transformação do uso do solo em dois setores distintos em Paragominas: a zona central do município onde predomina a agropecuária comercial e a zona de assentamentos e propriedades pequenas a Leste do município.

## 2. Metodologia

*Área de estudo*

O Município de Paragominas possui área de 19.330 km² e localiza-se a 320 km da capital do Pará Belém (Figura 1). Em 2010 apresentava uma população de 97.819 habitantes (IBGE, 2010). De acordo com dados do TerraClass de 2010, aproximadamente 73% da área do município era ocupada por florestas, vegetação secundária e reflorestamento, 21% por pastagens, 3,5% por agricultura comercial e 2,5 por outros usos.

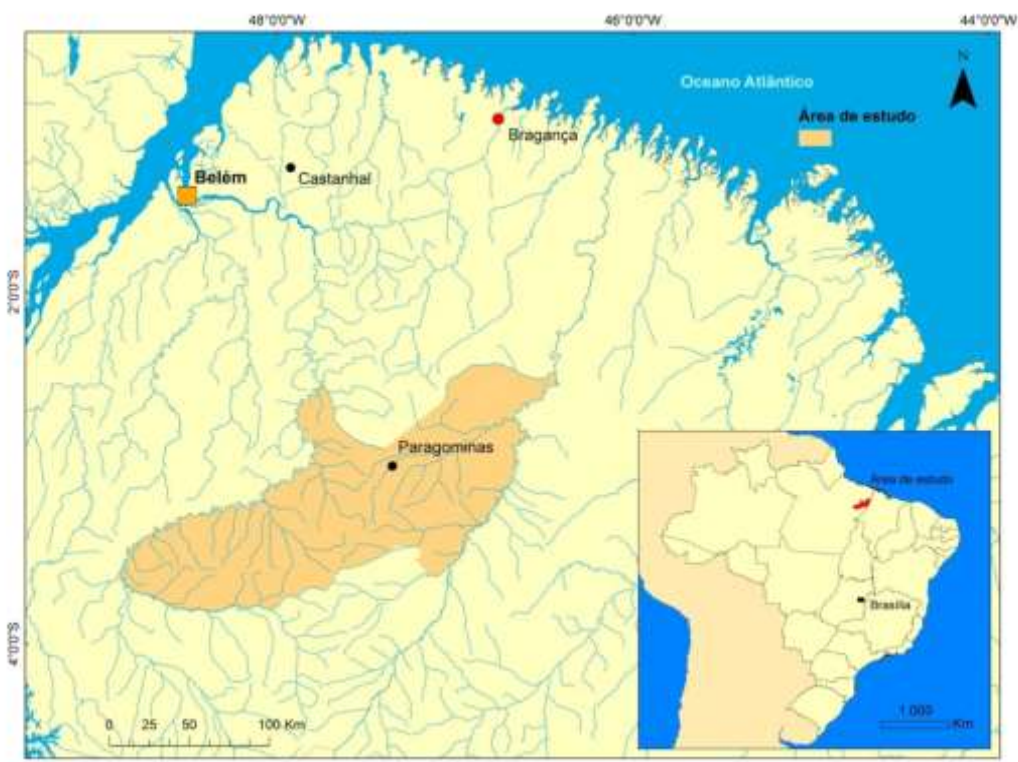

Figura 1: Localização do município de Paragominas no contexto da porção Leste do Pará e no contexto do Brasil.

Seu território está no contexto da bacia sedimentar do Grajaú, compreendendo arenitos caoliníticos da série Itapecurú e da formação Ipixuna, ambas recobertas por couraças lateríticas e por uma argila sedimentar denominada Argila de Belterra (KOTSCHOUBEY et al., 2005). Apresenta altitudes entre 160 e 190 m, e seu relevo é caracterizado por platôs (chamados localmente de "chapadas"), que são separados por vales que podem chegar a diversos quilômetros de largura, sendo que o contraste de altitude entre os platôs e os fundos de vale diminui do Sul em direção ao Norte (LAURENT et al., 2014).

*Tratamento dos dados*

Os testes de foram feitas no *software* DINAMICA EGO (SOARES-FILHO et al, 2002), que utiliza autômatos celulares para a simulação de mudanças de uso do solo. Os autômatos celulares operam de forma discreta no tempo e no espaço e permitem simplificar a modelagem de relações, fornecendo um ambiente estruturado onde vários níveis de interação e detalhe podem ser estudados. O mecanismo consiste em um arranjo de células no qual o estado de cada célula depende de seu estado prévio e de um conjunto de regras de transição, de acordo com o arranjo da vizinhança. Todas as células são atualizadas simultaneamente a passos discretos no tempo.

Esta técnica tem sido utilizada com sucesso em diversos casos de simulação de dinâmica de paisagem, em diferentes escalas espaciais, como em Linderman et al. (2004), Soares-Filho et al. (2002), entre outros.

O procedimento inicial consiste na montagem de um banco de imagens multitemporais e variáveis espaciais. Para o presente trabalho, foram utilizados os dados de uso do solo gerado por Perrier (2014) com dados do sensor MODIS para o município de Paragominas.

As variáveis espaciais consideradas foram: distância das rodovias principais, dimensão das propriedades, declividade, índice topográfico de umidade, textura do solo e distância dos rios. Após esta etapa inicial, foi feito o cálculo das matrizes de transição, que descrevem o percentual de mudança que um uso do solo no tempo t0 passa para outro tipo no tempo t1.

O passo seguinte foi o cálculo dos Pesos de Evidência, que são utilizados para produzir um mapa de probabilidades de transição. Este mapa espacializa as áreas com maior favorabilidade para determinada transição. O Peso de Evidência é um método bayesiano, onde o efeito de uma variável espacial sobre uma transição é calculado de forma independente de uma solução combinada (SOARES-FILHO et al., 2002). Os Pesos de Evidência representam a influência de cada variável sobre a probabilidade espacial da transição i-j (transição de um uso ao outro) e é calculada como segue:

$$O\{D|B\} = \frac{P\{D|B\}}{P\{\overline{D}|B\}} \quad (1)$$

$$\log\{D|B\} = \log\{D\} + W^+ \quad (2)$$

Onde W+ é o Peso de Evidência da ocorrência de um evento D, dado um padrão espacial B. A probabilidade posterior da transição i-j, dado um grupo de dados espaciais (B, C, D, ...N), é expressa como segue:

$$P\{i \Rightarrow j | B \cap C \cap D ... \cap N\} = \frac{e^{\sum W_N^+}}{1+e^{\sum W_N^+}} \tag{3}$$

Onde B, C, D e N são os valores de k variáveis espaciais que são mensuradas numa locação x,y e representada pelos seus pesos W+N. Como resultado é gerado um mapa de probabilidade de transição para cada tipo de transição que houver entre os períodos analisados.

A simulação é então feita utilizando-se o mapa de probabilidades de transição e as matrizes de transição. As mudanças são alocadas por meio de dois autômatos celulares complementares: *Expander*, responsável pela expansão ou contração de manchas prévias de determinada classe e *Patcher*, responsável pela formação de novas manchas (SOARES-FILHO et al., 2007). Estes operadores também devem ser parametrizados de acordo com as características da paisagem estudada: isometria, variância e tamanho médio das manchas de uso do solo.

Após a simulação, a validação foi feita pelo método de múltiplas janelas e função de decaimento constante, desenvolvida por Constanza (1989). Esta validação utiliza janelas de tamanhos crescentes para a comparação entre o mapa simulado e o mapa observado. Dentro de uma janela de dado tamanho, se o mesmo número de células das mesmas classes for encontrado, tanto na simulação quanto nos dados observados, a similaridade será 1. O resultado desta validação gera um gráfico que plota o tamanho da janela pela similaridade. O cálculo é feito dividindo-se o número total de pixels concordantes pelo número total de pixels observados e simulados (Figura 2).

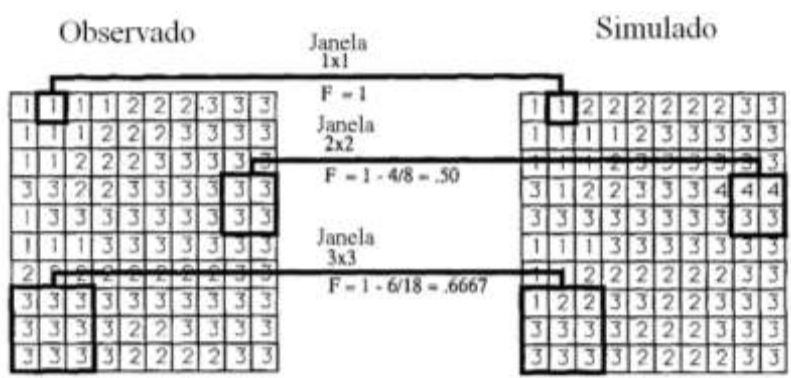

Figura 2: Exemplo do procedimento de validação por múltiplas janelas e decaimento constante. Fonte: Constanza (1989).

Para avaliar se os resultados de cada simulação foram satisfatórios, foi considerado o tamanho da janela necessária para se obter uma similaridade superior a 0,50.

## 3. Resultados/Discussões

*Uso do solo e variáveis espaciais*

Em um primeiro momento foram organizados os dados de entrada para a simulação. Os dados de uso do solo utilizados cenas de classificação de uso do solo baseadas no sensor MODIS dos anos 2004 e 2013 (Figura 3). Para uma análise dos anos intermediários, também foram utilizados dados dos anos 2007 e 2010.

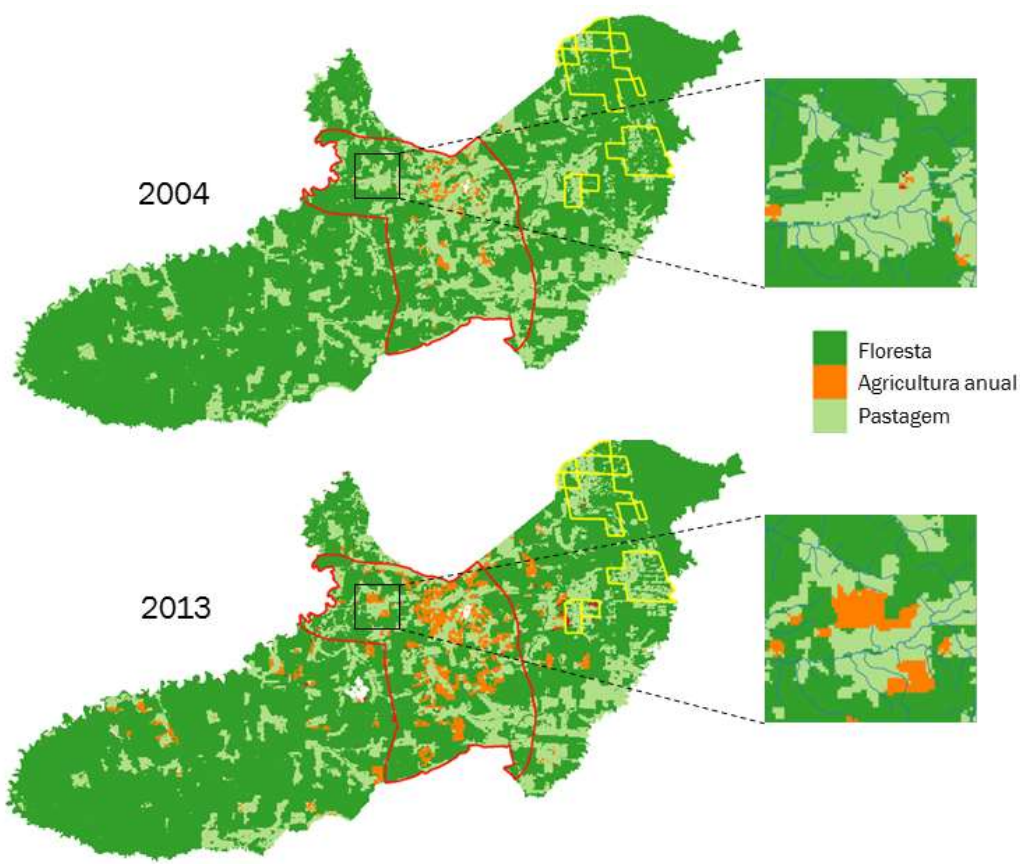

Figura 3: Classificações do uso do solo do município de Paragominas baseadas no sensor MODIS para os anos 2004 e 2013. Fonte: Perrier (2014). Delimitada em vermelho, a área de agropecuária comercial, e em amarelo, os assentamentos.

Com relação às variáveis espaciais, a textura dos solos foi umas das escolhidas pois orienta escolha de manejos na área de estudo, sobretudo a cultura de grãos, que é feita com preferência sobre a argila de Belterra. As zonas de "areia" como são conhecidas localmente (Areia argilosa), são ocupadas preferencialmente pela pecuária (Figura 4 – a). A declividade, por sua vez, possui um papel importante como limitante da mecanização da agricultura (Figura 4 – b).

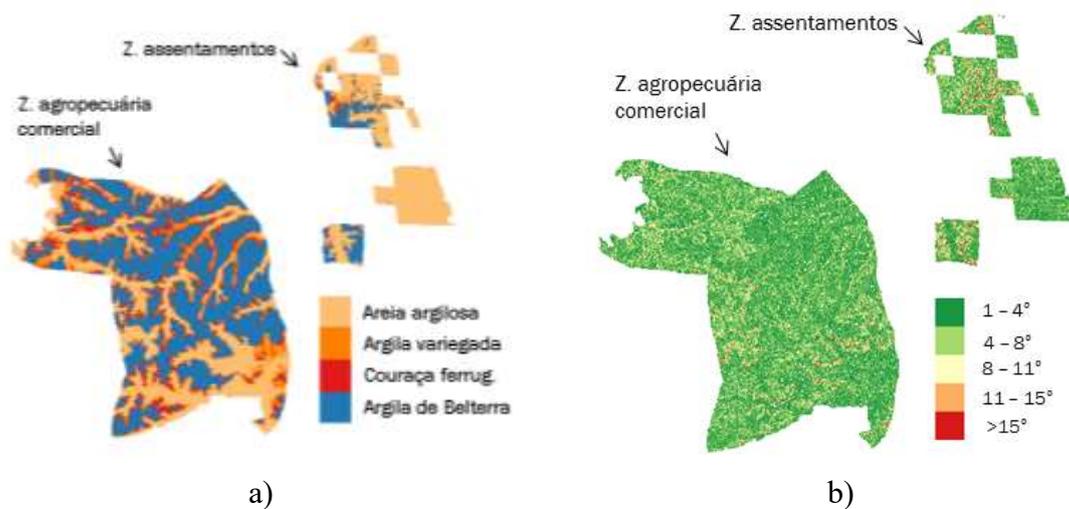

a)    b)

Figura 4: a) Textura do solo. Fonte: Laurent et al. (2014). b) Declividade em graus. Baseado em dados Topodata.

Os valores superiores do Índice Topográfico de Umidade predizem as áreas com maior probabilidade de ocorrência de solos saturados, fundos de vales e drenagens. Estas áreas são limitantes tanto para a agricultura quanto para a pecuária (Figura 5 – a). A distância de estradas em bom estado ou pavimentadas é uma variável reconhecida em diversos estudos de modelagem de mudança de uso do solo na Amazônia, associadas principalmente taxas de desmatamento (Figura 5 – b).

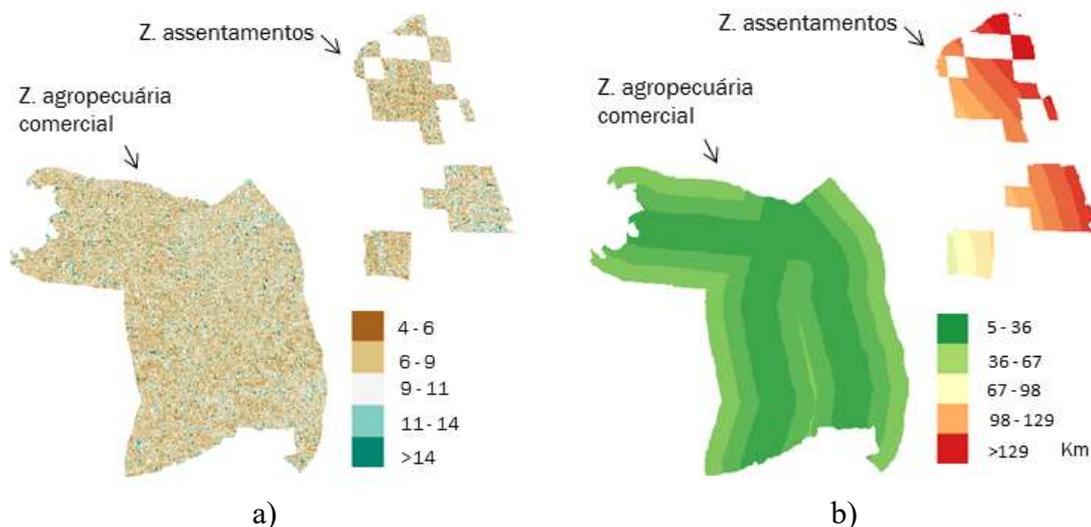

a)    b)

Figura 5: a) Índice Topográfico de Umidade. Baseado em dados Topodata. b) Distância de estradas em bom estado.

A dimensão das parcelas das propriedades representa o componente socieconômico e político que pode definir diferentes padrões de regras espaciais. Esta variável foi considerada somente para a zona central de agropecuária comercial, pois no contexto da zona de assentamentos, existe grande predominância de propriedades menores que 200 ha, que é a menor classe considerada (Figura 6 – a). Com relação à variável distância das principais drenagens, esta condiciona a aplicação da legislação ambiental (Áreas de Preservação Permanente), a presença de solos saturados e a disponibilidade de água para a dessedentação do gado (Figura 6 – b).

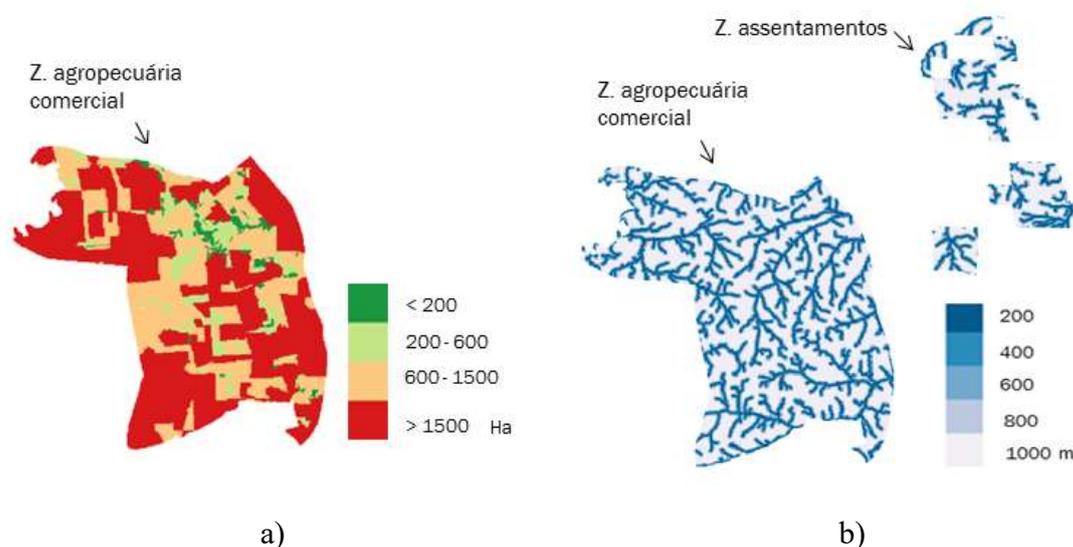

a)                                      b)

Figura 6: a) Tamanho das propriedades, em classes. b) Variável distância das drenagens. Elaborada com base em dados Topodata.

*Razões de transição*

A segunda etapa consiste na determinação das razões de transição. Estas razões indicam a proporção que determinado tipo de uso do solo se transformou em outro no período considerado (Tabela 1).

Tabela 1: Razões de transição da zona de agropecuária comercial para o período 2004-2013.

| Transições | Razões de transição |
|---|---|
| Floresta ➤ Pasto | 0,08 |
| Floresta ➤ Agricultura | 0,05 |
| Pasto ➤ Agricultura | 0,19 |
| Pasto ➤ Floresta | 0,16 |
| Agricultura ➤ Pasto | 0,14 |
| Agricultura ➤ Floresta | 0,02 |

Observa-se que a maior razão de transição para esta zona foi a de pasto para agricultura, o que reflete o processo de expansão do cultivo de grãos no município no período sobre as áreas de pastagens. O valor indica que aproximadamente 19% da área de pasto da zona definida foi transformada em agricultura no período considerado.

Para a zona de assentamentos, a primeira diferença é ausência da classe de uso do solo agricultura. Isso se deve ao fato de que a resolução espacial do sensor MODIS não é capaz de individualizar pequenos cultivos que porventura possam ocorrer nesta área. Aqueles existentes em geral acabam sendo interpretados na classificação como áreas de pastagem. Assim, para a área de assentamentos, a classe pastagem foi definida como pasto/agricultura (Tabela 2).

Tabela 2 : Razões de transição da zona de assentamentos para o período 2004-2013.

| Transições | Razões de transição |
|---|---|
| Flor➤Pasto/Agric | 0,35 |
| Pasto/Agric➤Flor | 0,09 |

Nas transições para a área de assentamentos observa-se aproximadamente 35% das áreas identificadas como floresta se transformaram em pasto/agricultura. A terceira etapa para a simulação é a determinação dos Pesos de Evidência, que são resultado da relação entre as transições e as variáveis espaciais.

*Simulação*

Na etapa de simulação da dinâmica do uso do solo no período 2004-2013, os operadores *Expander* e *Patcher* do DINAMICA EGO utilizam as razões de transição e os Pesos de Evidência observados para alocar as mudanças sobre o mapa de 2004 e assim criar uma simulação do ano 2013. Os resultados das simulações são demonstrados na Figura 7.

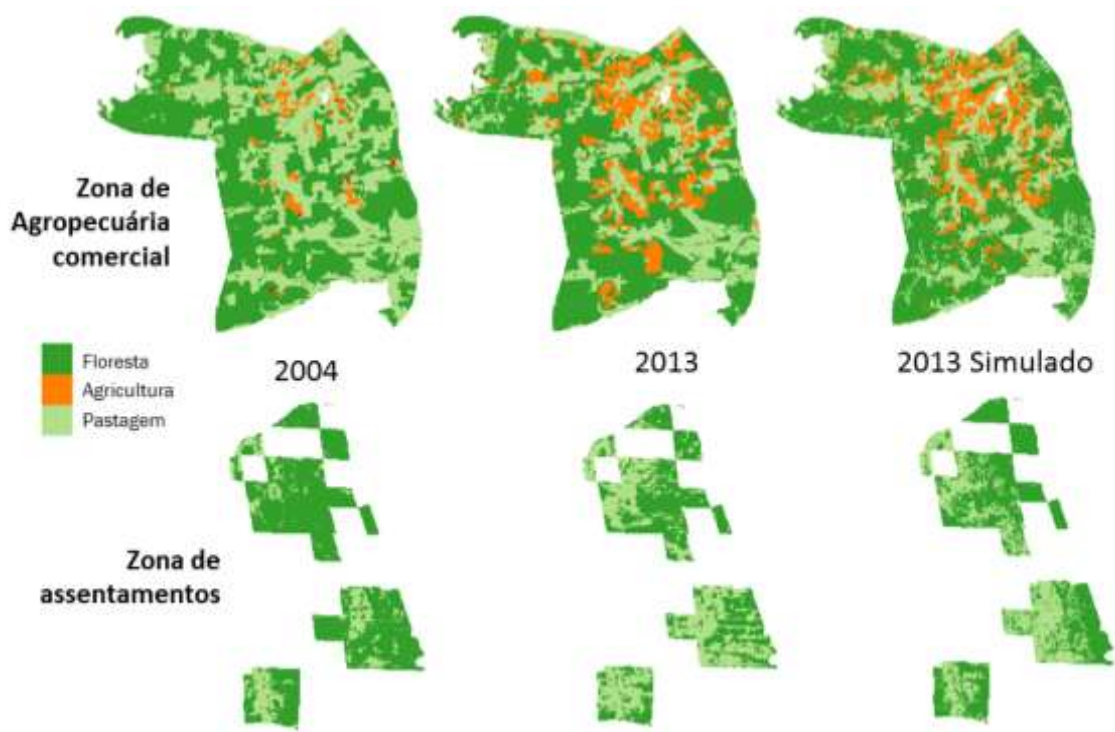

Figura 7: Dados de uso do solo observados dos anos 2004 e 2013 e simulações do ano 2013 das zonas estudadas.

*Validação*

Com relação à validação com o método de múltiplas janelas, para a simulação na zona de agropecuária comercial uma a similaridade igual a 0,5 é atingida com uma janela de 3 pixels (ou 750 m), indicando um bom resultado para a simulação. Para a zona de assentamentos, a similaridade a 0,5 foi atingida com uma janela de 1,5 pixels (ou 375 m), o que também sugere um bom resultado da predição do modelo.

*Regras espaciais*

Considerando que o modelo apresentou resultados satisfatórios para a simulação da mudança de uso do solo, assumiu-se que consequentemente as variáveis espaciais consideradas apresentam uma importância substancial no contexto da decisão dos produtores ao alocar as mudanças de uso do solo.

Algumas variáveis podem apresentar uma relação complexa, mas que pode auxiliar na interpretação dos condicionantes da mudança do uso do solo, como é o caso da textura do solo. Os gráficos da Figura 8 apresentam os Pesos de Evidência para a transição floresta ➤ pasto na zona de agropecuária comercial e floresta ➤ pasto/agricultura na zona de assentamentos para diferentes períodos entre 2004 e 2013.

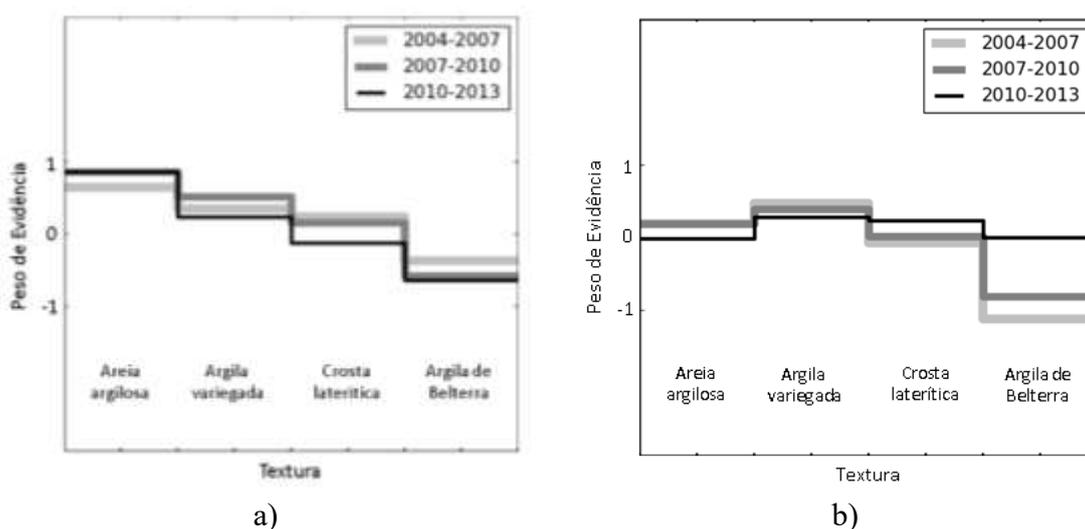

a)           b)

Figura 8: a) Pesos de Evidência da transição floresta ➤ pasto para diferentes texturas do solo na zona de agropecuária comercial. b) Pesos de Evidência da transição floresta ➤ pasto/agricultura para diferentes texturas do solo zona de assentamentos.

Observa-se que a relação da transição de uso do solo com as texturas do solo se faz de maneira diferente entra as zonas estudadas: na zona de agropecuária comercial o peso é positivo (favorável à transição) principalmente para a textura areia, enquanto que o peso é negativo (desfavorável à transição) para a argila (argila de Belterra).

Para a zona de assentamento, a textura com peso mais positivo para a transição floresta ➤ pasto/agricultura é a argila variegada, enquanto que o peso mais negativo também é sobre a argila de Belterra.

Nesse contexto cabe destacar que a textura do solo na área de estudo possui uma estreita relação com o relevo (LAURENT et al. 2014). As areias são comumente associadas aos fundos de vales, juntos dos rios e áreas com solos saturados. A argila variegada se situa em geral nas encostas, na transição entre os fundos de vale e as partes mais elevadas do relevo. Já a crosta laterítica e a argila de Belterra são associadas geralmente às partes mais elevadas do relevo (chapadas).

Com relação ao fato da maior parte do desmatamento (transição floresta ➤ pasto) na zona de agropecuária comercial estar relacionado à areia, isso levanta a seguinte hipótese: os desmatamentos que ocorreram no período estão relacionados às atividades pecuárias. A pecuária necessita de acesso facilitado às águas superficiais, como rios, para dessedentação do gado, enquanto que a baixa fertilidade das areias não afeta drasticamente a produção dos pastos da pecuária extensiva (PIKETTY et al., 2015). O desmatamento pelos fundos dos vales para expansão da pecuária é característica histórica da expansão da frente pioneira na área de estudo (ALMEIDA e UHL, 1998). Por outro lado, embora exista uma expansão do plantio de grãos sobre solos argilosos, que são mais exigentes com relação à qualidade do solo, esta se deu principalmente sobre áreas de pasto preexistente (vide Tabela 1), devido possivelmente à influência de ações como a implantação do programa Município Verde para coibir o desmatamento e a Moratória da Soja, mecanismo para evitar este plantio sobre áreas desmatadas (GIBS et al., 2015).

Considerando a variação temporal dos pesos, observa-se que no período 2004-2007 a amplitude dos mesmos era menor, tanto o positivo para a areia quanto o negativo para a argila. Para os anos seguintes, até 2013, esta amplitude aumenta. Isso também pode ser um reflexo das ações visando coibir o desmatamento, notadamente aqueles desmatamentos vinculados ao cultivo de grãos, pois se verifica que o peso negativo para desmatamento sobre argila fica mais negativo, indicando que menos desmatamentos foram realizados sobre argila nesta porção do município entre 2007 e 2013.

Na área dos assentamentos, é possível que o peso positivo para o desmatamento tenha se dado sobre a argila variegada devido ao histórico da expansão do desmatamento e à posição que os solos se colocam em relação a este histórico. Ao se observar o uso do solo dos assentamentos em 2004, nota-se que parte das áreas desmatadas se concentra junto das principais drenagens e fundos de vales, ou seja, junto de solos arenosos. Ao longo do tempo a expansão do desmatamento se deu em direção às encostas, onde predomina a argila variegada. Embora a argila variegada possua área bem inferior às outras texturas (vide Figura 3 – a), é possível que seja sobre esta que ocorreram a maior parte das transformações de floresta para pastou ou agricultura no período 2004-2013.

A dinâmica temporal dos pesos também parece contribuir para esta hipótese, pois no período inicial de 2004 a 2010 havia um peso maior para a argila variegada e um peso substancialmente mais negativo para a argila de Belterra, indicando que este período se caracterizou pelo franco desmatamento sobre as encostas, mas ainda com pouca expressão sobre o topo das chapadas, onde está a argila de Belterra.

Além disso, o do cultivo de pimenta-do-reino, que embora não pôde ser individualizado na classificação utilizada, pode ser uma causa da intensificação das transformações da floresta para outros usos nos assentamentos e a redução do peso negativo da transição floresta ➤ pasto/agricultura sobre a argila de Belterra no período 2010-2013. Este cultivo, que tem uma importância elevada no município, é realizado predominantemente por pequenos produtores e é mais bem adaptado aos solos argilosos (SANTOS et al., 2015), como a argila de Belterra, presentes nas chapadas que ainda possuem grande proporção de florestas na zona de assentamentos.

## 4. Considerações Finais

O presente trabalho apresentou os resultados da análise da mudança do uso do solo e a relação com variáveis espaciais em duas zonas distintas no município de Paragominas-PA. Com as informações obtidas, modelos foram elaborados e simulações de mudança do uso do solo foram feitas, obtendo-se resultados satisfatórios nas validações, indicando que as variáveis espaciais textura do solo, distância das estradas, distância dos rios, declividade, índice topográfico de umidade e dimensão das propriedades são capazes de predizer boa parte da dinâmica do uso do solo, tanto na zona de agropecuária comercial quanto na zona de assentamentos e propriedades pequenas.

Entre as zonas analisadas também foram observadas regras espaciais de ocupação diferentes. Com relação ao solo, o exemplo discutido, colocou-se a hipótese de que este condiciona parte da decisão de mudança de uso do solo na zona de agropecuária comercial principalmente devido a relação da pecuária com as áreas arenosas de fundo de vale e a valorização da argila de Belterra para o cultivo de grãos nas chapadas.

Nos assentamentos, por outro lado, a dinâmica observada pode estar relacionada mais com o histórico de ocupação dos assentamentos do que com a qualidade dos solos, embora possa haver também uma valorização da argila para o cultivo de pimenta-do-reino.

Estas simulações poderão ser aprimoradas no futuro com o uso de variáveis espaciais e classificações de uso do solo mais precisas, bem como informações obtidas em campo, o que permitirá identificar áreas com maior susceptibilidade de expansão de determinados usos e auxiliar projetos de planejamento territorial.

## 5. Agradecimentos

## 6. Referências Bibliográficas

ALMEIDA, O.T.; UHL, C. **Planejamento do uso do solo do Município de Paragominas Utilizando dados Econômicos e Ecológicos**. 1º ed., IMAZON. Belém, Brazil, 1998.

CONSTANZA, R. Model goodness of fit: a multiple resolution procedure. **Ecological Modelling**, v.47, n.1. p.199-215, 1989.

GIBBS, B. H. K.; RAUSCH, L.; MUNGER, J.; et al. Brazil's Soy Moratorium. **Science**, v. 347, n. 6220, p. 377–378, 2015.

IBGE. **Censo 2010**. Disponível em:http://www. censo2010. ibge. gov. br/.

KOTSCHOUBEY, B., CALAF, J. M. C., COSTA LOBATO, A. C., SABA LEITE, A., DUARTE AZEVEDO, C. H. Caracterização e gênese dos depósitos de bauxita da província bauxitífera de Paragominas, Noroeste da Bacia do Grajaú, Nordeste do Pará/Oeste do Maranhão. In. MARINI, O. J. et al. **Caracterização de depósitos minerais em distritos mineiros da Amazônia**. Brasília, ADIMB (Agência para o Desenvolvimento Tecnológico da Indústria Mineral Brasileira), 2005.

LAURENT, F.; POCCARD-CHAPUIS, R.; PLASSIN, S. 2014. Cartographie de la texture des sols a partir du relief en amazonie orientale. **Environnement et géomatique: approches comparées France-Brésil**. Rennes, 2014.

LINDERMAN, M. A.; AN, L.; BEARER, S.; HE, G.; OUYANG, Z.; LIU, J. Modeling the spatio-temporal dynamics and interactions of households, landscapes, and giant panda habitat. **Ecological Modelling**, nº 183, v. 1, p. 47–65, 2004.

PERRIER, F. Caractérisation et cartographie de l'écoefficience des pâturages amazoniens à l'aide d'images MODIS. **Mémoire de recherche**. Université du Maine, 2014.

PERZ, S.; MESSINA, J. P.; REIS, E.; WALKER, R.; WALSH, S. J. Cenários futuros de paisagens amazônicas: modelos econométricos e de simulação de dinâmica (Original em Inglês). In: KELLER, M. et al. **Amazonia and Global Change**. American Geophysical Union, 2009.

PIKETTY, M. et al. Multi-level governance of land use changes in the brazilian amazon: lessons from Paragominas, state of Pará. **Forests**, n. 6, p. 1516-1536, 2015.

PINTO, A.; AMARAL, P.; SOUZA JUNIOR, C.; et al. **Diagnóstico socioeconômico e florestal do município de Paragominas**. IMAZON, 2009.

SANTOS, C.A; ALVES, L.F.N.; FARIAS, M.H.C.S; PINHEIRO, J.B.S. Estudo preliminar do processo de introdução de novas atividades produtivas: o caso da pimenta-do-reino no assentamento Luiz Inácio em Paragominas – PA. Geoambiente, n. 25. Jul.-Dez. p. 1–13, 2015.

SOARES-FILHO, B. S.; CERQUEIRA, G. C.; PENNACHIN, C. L.. DINAMICA - A stochastic cellular automata model designed to simulate the landscape dynamics in an Amazonian colonization frontier. **Ecological Modelling**, n. 154, v. 3, 217–235, 2002.

SOARES-FILHO, B.S. et al. Modelagem de dinâmica da paisagem: concepção e potencial de aplicação de modelos de simulação baseados em autômato celular. In.: SILVA, J.M.C. et al. **Megadiversidade: Modelagem ambiental e conservação da biodiversidade.** Conservação Internacional, v.3, 2007.